\documentstyle[aps,prl,epsfig,multicol]{revtex}

\input epsf
\begin{document}

\newcommand \be {\begin{equation}}
\newcommand \ee {\end{equation}}
\newcommand \bea {\begin{eqnarray}}
\newcommand \eea {\end{eqnarray}}
\newcommand \nn {\nonumber}

\title{Effective hyperfine temperature in frustrated Gd$_2$Sn$_2$O$_7$: \\
two level model and $^{155}$Gd M\"ossbauer measurements} 

\author{E. Bertin, P. Bonville, J.-P. Bouchaud, J.A. Hodges}
\address{\it Commissariat \`a l'Energie Atomique,
Service de Physique de l'\'Etat Condens\'e\\
91191 Gif-sur-Yvette, France}
\author{J.P. Sanchez, P. Vulliet}
\address{\it Commissariat \`a l'Energie Atomique, Service de Physique
Statistique, Magn\'etisme et Supraconductivit\'e\\
38054 Grenoble, France}

\maketitle

\begin{abstract}
Using $^{155}$Gd M\"ossbauer spectroscopy down to 27\,mK, we show that, in the 
geometrically frustrated pyrochlore Gd$_2$Sn$_2$O$_7$, the Gd$^{3+}$ 
hyperfine levels are populated out of equilibrium. From this, we deduce that
the hyperfine field, and the correlated Gd$^{3+}$ moments which produce
this field, continue to fluctuate as T $\to$ 0.
With a model of a spin 1/2 system experiencing a magnetic field
which reverses randomly in time, we obtain an analytical expression for the
stationary probability distribution of the level populations.
This distribution is a simple function of the ratio of the
nuclear spin relaxation time to the average electronic spin-flip time. 
In Gd$_2$Sn$_2$O$_7$, we find the two time scales are of the same order of 
magnitude.
We discuss the mechanism giving rise to the nuclear spin relaxation and 
the influence of the electronic spin fluctuations on the hyperfine specific 
heat. The corresponding low temperature measurements in Gd$_2$Ti$_2$O$_7$ are 
presented and discussed.

\end{abstract}

\begin{multicols}{2}
\section {Introduction} \label{intr}

In geometrically frustrated magnetic systems,  the interaction
energies cannot be minimised at the same time for all pairs of spins 
\cite{diep}. A simple example is the two-dimensional triangular
lattice with nearest-neighbour antiferromagnetic (AF) coupling of isotropic 
spins (the so-called Heisenberg antiferromagnet). 
An analogous situation in three dimensions is provided by the pyrochlore
structure compounds with general formula R$_2$M$_2$O$_7$ where 
the rare earth (R) ions and the transition metal or $sp$-metal (M) 
ions each lie on an array of corner-sharing tetrahedra.
A general prediction was made by Villain \cite{vill} about the ground state of
a Heisenberg antiferromagnet on a tetrahedral lattice: as the temperature goes 
to zero, there is no long range ordering and the system remains in a 
collective paramagnetic state where spin fluctuations persist. 
The low temperature properties in frustrated systems may also be influenced by 
perturbations beyond the dominant nearest neighbour exchange interaction
\cite{moessner98a}. These include exchange with more distant neighbours 
\cite{reimers91a,reimers91b,kinney79}, dipolar coupling \cite{palmer99},
and anisotropy \cite{bramwell94,moessner98b}.
They may lead to the lifting of the degeneracy
of the ground state and to the development of magnetic long range order. /par

The gadolinium based pyrochlore compounds Gd$_2$Ti$_2$O$_7$ and
Gd$_2$Sn$_2$O$_7$, where the M sublattice is non-magnetic, possess the basic
ingredients of a Heisenberg antiferromagnet.
The Gd$^{3+}$ ion has zero orbital moment ($L$=0, $S$=7/2) and thus presents
a (quasi) isotropic reponse to an exchange or external field.  Both compounds 
show a negative paramagnetic Curie-Weiss temperature ($\theta_P=-$10\,K for 
Gd$_2$Ti$_2$O$_7$ \cite{raju} and $-$8\,K for Gd$_2$Sn$_2$O$_7$ \cite{hodg}) 
indicative of AF coupling.
Experimentally, however, it is known that the properties of the Gd$^{3+}$ 
based 
pyrochlores do show some differences relative to those expected for the
nearest neighbour AF Heisenberg model. For example, in Gd$_2$Ti$_2$O$_7$, 
specific heat data suggest that a phase transition occurs near 1\,K \cite{raju}
and below this temperature neutron diffraction data evidence
magnetic Bragg peaks \cite{champion}.
The details of the low temperature properties of the Gd$^{3+}$ based 
pyrochlores are however quite complex. 
We find that in both Gd$_2$Sn$_2$O$_7$ and Gd$_2$Ti$_2$O$_7$
there are two closely separated magnetic transitions near 1\,K   
\cite{hodg} and we show here that, in Gd$_2$Sn$_2$O$_7$, spin fluctuations
clearly continue to exist as T $\to$ 0. 

The evidence for the very low temperature spin fluctuations is provided
by $^{155}$Gd M\"ossbauer measurements, but not through 
the conventional approach based on the analysis of the M\"ossbauer line 
shape to obtain the relaxation rate. This approach is possible only when the 
fluctuation rate
of the spins falls within the 
classical M\"ossbauer ``relaxation window'', which is centered 
around 100\,MHz for $^{155}$Gd. In the present case, the spins fluctuate at 
lower frequencies 
such that the hyperfine field associated with the Gd$^{3+}$ spin appears 
static on the M\"ossbauer time scale.
The very low temperature spin fluctuations were evidenced using an original 
method, i.e. through the observation that the hyperfine levels of 
the $^{155}$Gd nuclei are populated out of thermal equilibrium. 
We show that an out-of-equilibrium distribution can occur when the electronic
spin flips persist at low temperature and when the nuclear
relaxation time $T_1$ is longer than or of the same magnitude as the flipping
time $\tau$ of the hyperfine field (of the electronic spin).
Considering the nuclear spins as a two-level system driven by a randomly
fluctuating field, we develop a stochastic model which yields an 
analytical expression for the probability distribution of the level 
populations.
This quantity depends on the ratio $T_1/\tau$ of the two characteristic times 
of the system and it is directly linked to the effective hyperfine temperature
provided by the very low temperature M\"ossbauer measurements.

We also propose a nuclear relaxation mechanism, linked with the scattering of
electronic spin-waves, which could explain short nuclear relaxation times
suggested by the analysis and we discuss some implications of the presence of 
spin fluctuations on the magnitude of the hyperfine specific heat.

\section{The low temperature $^{155}$Gd M\"ossbauer measurements} \label{moss}

The pyrochlore lattice corresponds to the cubic space group $Fd3m$ and
the crystallographic unit cell contains 16 Gd$^{3+}$ ions each
with the same point symmetry ${\bar 3}m$. Each Gd$^{3+}$ ion has one of the 
four [111] directions as threefold symmetry axis. 
The M\"ossbauer
transition of the $^{155}$Gd isotope links the ground nuclear state, with spin
$I_g$=3/2 to the first excited state, with spin $I_e$=5/2, at an energy
$E_0$=86.5\,keV. The M\"ossbauer spectra were recorded in the range from 
4.2\,K down to 27\,mK, in a $^3$He-$^4$He dilution refrigerator coupled to a 
constant acceleration spectrometer and using a Sm($^{155}$Eu)Pd$_3$ source.
For $^{155}$Gd, the velocity unit 1\,mm/s corresponds to 69.8\,MHz or 
to 3.35\,mK. 

At 4.2\,K, in agreement with the literature
\cite{cashion73,armon_nucl}, for each of the two compounds we observe a
quadrupole hyperfine spectrum characteristic of the paramagnetic phase.
The quadrupole splitting is $-$8 and $-$11\,mm/s in 
Gd$_2$Sn$_2$O$_7$ and in Gd$_2$Ti$_2$O$_7$ respectively, corresponding
respectively to a 
splitting of 13.5 and 18.4\,mK between the $m=\pm 3/2$ and 
$m=\pm 1/2$ sublevels of the $I_g$=3/2 $^{155}$Gd nuclear ground state.
As Gd$^{3+}$ is an S-state ion, in these insulating pyrochlores, the hyperfine
quadrupole interaction is due only to
the electric field gradient produced by the anisotropic  distribution of
lattice charges. 

Below $\sim$ 1\,K, an additional magnetic hyperfine interaction is
visible. Usually, the presence of a hyperfine field
is linked with magnetic ordering, but short range correlated moments
also yield a hyperfine field spectrum provided their fluctuation frequency is 
lower than a characteristic value which is about 100\,MHz for $^{155}$Gd. 
The thermal evolution of the magnetic hyperfine parameters and the
information they provide concerning low temperature magnetic properties 
will be discussed in a separate publication \cite{hodg}. 
Let us only remark here that 
in each compound there are two closely separated magnetic transitions near 
1\,K and in the low temperature saturated state, each compound shows a 
unique hyperfine field (a unique size for the spontaneous Gd$^{3+}$ moment) 
which is oriented perpendicular to its associated [111] axis. Our finding that
all the Gd$^{3+}$ ions carry the same sized moment perpendicular to the
symmetry axis [111] contrasts with the magnetic structure proposed for 
Gd$_2$Ti$_2$O$_7$ in 
Ref.\onlinecite{champion}, where one out of four Gd$^{3+}$ ions is claimed to
have zero moment. The two sets of results could be reconciled only with the
unlikely assumption that one of the four Gd$^{3+}$ moments in each tetrahedron,
perpendicular to the [111] axis, shows no long range correlations.

\begin{figure}
\centerline{
\epsfxsize = 8.5cm
\epsfbox{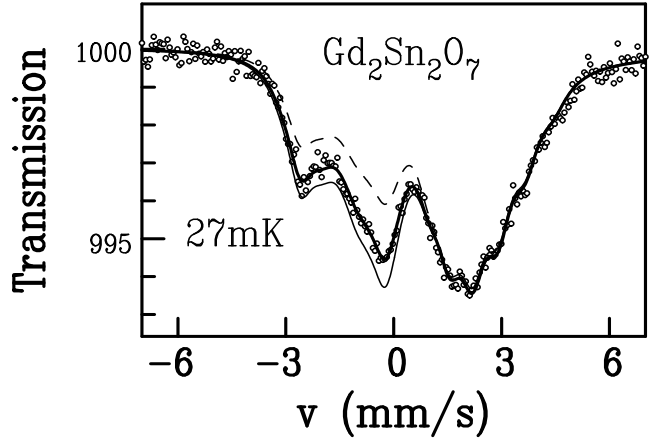}}
\vskip 0.5 cm
\caption{\sl $^{155}$Gd M\"ossbauer absorption spectrum at 27\,mK in 
Gd$_2$Sn$_2$O$_7$. The sample temperature is at most 33(2)\,mK (see text).
The dashed line represents the expected theoretical 
spectrum at $T$=27\,mK, the thin solid line the expected spectrum for
equipopulated hyperfine levels ($T>0.2$\,K) and the thick solid line passing
through the experimental points the 
spectrum with a fitted effective temperature $T_{eff}$=90\,mK.}
\label{spe3t}
\end{figure}

\begin{figure}
\centerline{
\epsfxsize = 8.5cm
\epsfbox{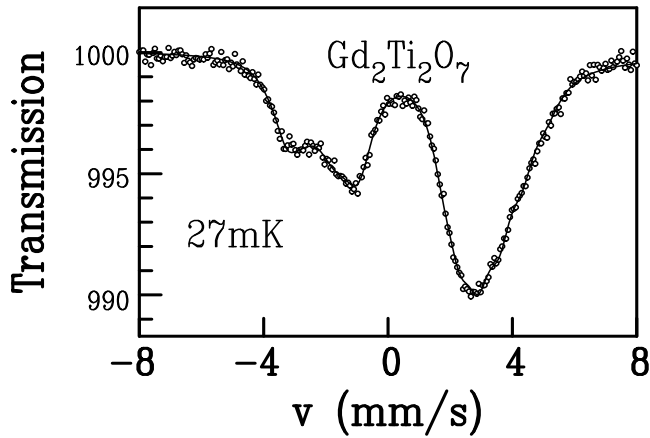}}
\vskip 0.5 cm
\caption{\sl $^{155}$Gd M\"ossbauer absorption spectrum at 27\,mK in 
Gd$_2$Ti$_2$O$_7$. The sample temperature is at most 33(2)\,mK (see text).
The fitted line was obtained with an effective
temperature of 36\,mK.}
\label{figti}
\end{figure}

In the following, we will focus on the data obtained at 27\,mK, shown in 
Fig.\ref{spe3t} for Gd$_2$Sn$_2$O$_7$ and in Fig.\ref{figti} for 
Gd$_2$Ti$_2$O$_7$.
For Gd$_2$Sn$_2$O$_7$, the fitted hyperfine field is 30\,T,
and the combined quadrupolar and magnetic interactions within the ground 
$I_g$=3/2 
nuclear state yield 4 hyperfine levels with energies 0, 0.05, 12.1 and 
15.9\,mK. Therefore, at a temperature not too high with respect to these
splittings, the intensities of the M\"ossbauer absorption transitions
originating from these levels should reflect their different populations.
In other words, the effective hyperfine temperature can be obtained from 
the line intensities of the $^{155}$Gd M\"ossbauer spectrum insofar that 
the temperature is below about 100\,mK. The 27\,mK spectrum in 
Gd$_2$Sn$_2$O$_7$ of 
Fig.\ref{spe3t} was fitted with the temperature as a free parameter. 
Surprisingly, the best fit, shown as a thick solid line in Fig.\ref{spe3t},
yields a temperature of 90\,mK.
Taking into account experimental uncertainties, the hyperfine level
temperature falls in the range from 60 to 120\,mK. Also represented in 
Fig.\ref{spe3t} are the expected spectrum for $T$=27\,mK (dashed line) and
the limiting high temperature ($T>200$\,mK) spectrum when the 
hyperfine levels are equi-populated (thin solid line). Thus, it is clear
the difference between the measured effective hyperfine temperature 
and the sample temperature is outside statistical errors. 
In this analysis, lineshape effects that could lead to intensity
deviations were fully accounted for:
the dispersive correction \cite{czjzek} was included with $\xi$=0.027, as
was the Goldanskii-Karyagin effect shown to be present in the Gd
pyrochlores \cite{armon74} with an anisotropy coefficient $\epsilon=-$1.5.

We checked that the M\"ossbauer absorber, which is in the form of a powder
mixed with General Electric varnish and glued onto a thin copper sheet,
was correctly thermalised by carrying out two controls.
First we performed, a 27\,mK
measurement with the isotope $^{151}$Eu (I$_g$=5/2, I$_e$=7/2, 
$E_0$=21.6\,keV) in the
insulating compound EuAl$_2$Si$_2$O$_8$ (Eu$^{2+}$ charge state)
in its saturated magnetically ordered state. 
As the hyperfine interaction is larger for $^{151}$Eu$^{2+}$ than for 
$^{155}$Gd$^{3+}$, Eu$^{2+}$ compounds are better ``M\"ossbauer thermometers'' 
\cite{shenoy} than are Gd$^{3+}$ compounds. From these measurements, we found 
that the $^{151}$Eu hyperfine level temperature corresponds exactly
to the sample temperature. 
Second, we performed, under identical experimental conditions as for the 
pyrochlores, a 27\,mK $^{155}$Gd M\"ossbauer measurement in the metallic
Gd-based ferromagnet GdCo$_2$B$_2$ ($T_C$=26\,K). This compound was selected
because the size of the quadrupole hyperfine interaction and the size and
direction of the saturated hyperfine field relative to the principal axis of 
the electric field gradient \cite{felner84} are close to those in the 
pyrochlores. 

We obtained a hyperfine level temperature of 33(2)\,mK, slightly 
higher than that of the sample. This difference could be due to the heating 
of the sample by the incident high energy $\gamma$-rays, so that a small 
temperature 
gradient exists between the sample and the carbon temperature probe. 
Such heating would also occur for the $^{155}$Gd measuremments in the 
pyrochlores meaning that for a temperature probe value of 27\,mK, the actual 
sample temperature could be 33(2)\,mK. We thus reach the conclusion that 
in Gd$_2$Sn$_2$O$_7$, the sample temperature is at most 35\,mK, but 
that the hyperfine levels have an effective temperature of 
90(30)\,mK; they are thus populated out of thermal equilibrium. 

From the $^{155}$Gd M\"ossbauer analysis of the 27\,mK data in 
Gd$_2$Ti$_2$O$_7$ (Fig.\ref{figti}), we find that the hyperfine field is
28.3\,T, yielding four hyperfine sublevels situated at 0, 0.02, 17.0
and 20.4\,mK. In contrast to Gd$_2$Sn$_2$O$_7$,
the effective hyperfine temperature is found to be 36\,mK. 
As this is only very marginally higher than the sample temperature (33(2)\,mK),
there is no clear experimental evidence that the hyperfine 
level temperature is different from that of the sample. The hyperfine levels 
can be said to be practically in thermal equilibrium. 
As the measurements in Gd$_2$Sn$_2$O$_7$ and in Gd$_2$Ti$_2$O$_7$ were carried 
out under exactly the same experimental conditions,
the fact that there is at most only a small difference in the two 
temperatures for Gd$_2$Ti$_2$O$_7$, whereas there is a marked difference
for Gd$_2$Sn$_2$O$_7$ reinforces the credibilty of the anomalous hyperfine 
level temperature found for Gd$_2$Sn$_2$O$_7$.

Our interpretation of the non-Boltzmann population distribution in 
Gd$_2$Sn$_2$O$_7$ is based on the influence of spin dynamics. The hyperfine 
levels
do not reach thermal equilibrium because the Gd$^{3+}$ hyperfine field
continues to
fluctuate as T $\to$ 0. In turn, this means the Gd$^{3+}$ magnetic moments 
which are at the origin of the hyperfine field continue to fluctuate as 
T $\to$ 0. As shown in section \ref{hypeff}, this approach also entails that 
the hyperfine spin-lattice times are of the same magnitude as those associated
with the fluctuations of the Gd$^{3+}$ hyperfine field.

As described above, the quadrupole and magnetic hyperfine interactions in 
Gd$_2$Sn$_2$O$_7$ lift the degeneracy of the $I_g$=3/2 ground state to leave 
two almost degenerate levels at 0 and 0.05\,mK and two closely separated 
levels at 12.1 and 15.9\,mK.  
To a first approximation, this resembles a two level system with an energy 
separation $\sim$ 14\,mK.
In the next section we present a model of a spin 1/2
system driven by a randomly varying (hyperfine) field. Although the 
$^{155}$Gd mixed nuclear 
levels in Gd$_2$Sn$_2$O$_7$ cannot be mapped exactly onto a spin 1/2 system,
the main results of our calculation are not affected by the details
of the nuclear wave-functions.\par

\section{The model of a two-level system driven by a randomly varying
field} \label{model}

In order to put on a more quantitative ground the concept of effective
temperature due to the competition between nuclear relaxation with time
scale $T_1$ and electronic spin flip, with time scale $\tau$, we
performed a model calculation on a two-level (nuclear) spin 1/2 system 
driven by a randomly time dependent (hyperfine) field. 

We wish to calculate the stationary out of equilibrium distributions of the
populations of the two levels.
We show here that the problem of a two level system in a randomly flipping 
field is exactly soluble, and that the stationary probability distribution can
be expressed in terms of the ratio $\mu=T_1/\tau$.
A two level system need not necessarily be a spin 1/2 system. The low 
temperature 
properties of glasses, for example, are argued to be dominated by quasi 
degenerate local configurations, where an atom or molecule hops between
two local equilibrium positions \cite{AHV,EntDistrib}. The 
difference of energies between these
two positions plays the role of the Zeeman splitting for a spin 1/2. If these 
two level
systems are strongly interacting, then the splitting field itself will have 
non trivial dynamics, and our calculation might be relevant to this 
situation as well.\par

Our calculation provides an exact solution for
the stationary distribution of a non equilibrium, randomly driven system.  
For a two level system, it is justified to describe the relaxation dynamics in
terms of a single relaxation time $T_1$. This
comes from the fact that the Master evolution operator for the `up' 
probability $P_u(t)$ and 
`down' probability $P_d(t)$  has two eigenvalues -- 
one is zero and corresponds to the 
Boltzmann equilibrium, and the second is $1/T_1$. 
Here we will consider a splitting field $H(t)=\pm H_0$, 
which changes its direction (or sign) randomly in time. We assume that the
probability of switching per unit time is a constant $1/\tau$. Then the time 
interval $t_f$ between two 
successive flips is distributed according to an exponential law of mean $\tau$:
\be
\rho(t_f)=\frac{1}{\tau}\, e^{-t_f/\tau}.
\ee
We are interested in the distributions of the populations of the two levels, 
but it turns out
that the calculation is easier if we use the magnetization $M(t)$:
\be
M(t) = m_0[P_u(t) - P_d(t)],
\ee
where $\pm m_0$ is the intrinsic moment of the spin levels. This is due to
the symmetry properties of the corresponding distributions (see
Eqn.\ref{symm} below).
Within the time interval during which $H(t)$ is constant, 
we can give an explicit expression for $M(t)$, that we denote by $M^+(t)$ 
or $M^-(t)$, depending on the sign of $H(t)$. 
Suppose $H(t)$ is positive between $t_1$ and $t_2$, and then negative 
between $t_2$ and $t_3$. Then we can write, for $t_1 \leq t \leq t_2$:
\be
M^+(t)=M_0\, (1-e^{-(t-t_1)/T_1}) + M^-(t_1)\, e^{-(t-t_1)/T_1}
\ee
and for $t_2 \leq t \leq t_3$:
\be
M^-(t)=-M_0\, (1-e^{-(t-t_2)/T_1}) + M^+(t_2)\, e^{-(t-t_2)/T_1}.
\ee
where $M_0=m_0 \tanh{{m_0 H_0} \over {k_B T}}$ is the Boltzmann magnetisation
of the system under a static magnetic field $H_0$.
Since the times $t_i$ are randomly distributed, we 
must determine the sequence of random magnetizations corresponding to 
the flipping times, $M^-(t_{2i+1})$ and $M^+(t_{2i+2})$. In the
stationary state, these quantities are identically distributed random 
variables $M^-$ and $M^+$, with some probability distributions 
${\cal P}_-(M^{-})$ and ${\cal P}_+(M^{+})$. 
Between two successive flips, $M^{+}$ and $M^{-}$ are related by:
\be
M^{+} = M_0 (1-e^{-t_f/T_1}) + M^{-} e^{-t_f/T_1},
\ee
and vice versa.
Obviously, both $M^+$ and $M^-$ will be in the interval $[-M_0,M_0]$. 
The stationarity of the probability distributions allows us to write the 
following equation:
\bea
{\cal P}_+&(&M^{+})= \int_{-M_0}^{M_0} 
dM^{-}\, {\cal P}_-(M^{-})\, \int_0^{+\infty} dt_f\, \rho(t_f) \: \times \nn \\
&\delta& \left[M^{+}-\left(M_0(1-e^{-t_f/T_1})+
M^{-}e^{-t_f/T_1}\right)\right] .
\eea
For reasons of symmetry, ${\cal P}_+$ and ${\cal P}_-$ satisfy 
the following relation:
\be
{\cal P}_-(M) = {\cal P}_+(-M),
\label{symm}
\ee
so that we get an integral equation involving only one distribution. It will 
be useful to make the following changes of variables:
\be
u=e^{-t_f/T_1},\qquad
y=\frac{M^{+}}{M_0},\qquad
z=\frac{M^{-}}{M_0}.
\ee
One therefore also has: $P_+(y)=M_0 \, {\cal P}_+(M^{+})$.
The resulting integral equation is:
\bea
P_+(y)= \int_{-1}^1 dz\, P_+(-z&)&\, \int_0^1 \frac{T_1\,du}{u}\, \times \nn \\
\frac{u^
{T_1/\tau}}{\tau}\, &\delta& [y-((1-u)+zu)].
\eea
Introducing the ratio $\mu=T_1/\tau$ of the relaxation time to the 
flipping time and using the properties of the $\delta$ distribution we find:

\be
P_+(y)= \int_{-1}^1 dz\, \frac{P_+(-z)}{1-z} \int_0^1 \mu\,du\,u^{\mu-1}\, 
\delta\left(u-\frac{1-y}{1-z}\right).
\ee
For the integral over $u$ to be non-zero, we must have $\frac{1-y}{1-z}<1$, or 
equivalently $z<y$. Changing $z$ in $-z$, we obtain the resulting integral 
equation:
\be
P_+(y)=\mu (1-y)^{\mu-1} \int_{-y}^1 \frac{P_+(z)}{(1+z)^\mu} dz.
\ee
It is easily checked that the following beta-distribution is an exact solution 
of this integral equation:
\be
P_+(y) = \frac{\Gamma(\frac{1}{2}+\mu)}{\Gamma(\frac{1}{2})\,\Gamma(\mu)}\, 
(1-y)^{\mu-1} (1+y)^\mu,
\label{pplus}
\ee
where $\Gamma(x)$ is the usual Gamma function defined as:
$\Gamma(x) = \int_0^{+\infty} t^{x-1}e^{-t} dt$.
Now we are interested in the distribution of magnetization $M(t)$ at 
{\it any} time $t$, not necessarily a flipping time. However, since the
flip times are chosen at random,
the distribution we have just computed is also that governing the 
magnetization observed at an arbitrary instant of time. \par

This distribution $P_+(y)$ must be used to compute the average populations
of the energy levels, as will be done in the next section. Here, we just
derive the mean distribution $P(y)$ of the (rescaled) magnetization, obtained 
by averaging over the periods of positive and negative $H$. 
Since the two values of $H$ are equally probable, we can write $P(y)$ as:
\bea
P(y)&=&\frac{1}{2} \left[P_+(y)+P_+(-y)\right]\\
&=&\frac{\Gamma(\frac{1}{2}+\mu)}{\Gamma(\frac{1}{2})\,\Gamma(\mu)}\, 
(1-y^2)^{\mu-1},
\eea
which is of course an even expression in $y$, leading to a zero mean value
for the magnetization. For $\mu \to 0$, this distribution
tends to two $\delta$ functions at $\pm 1$, as it should. Interestingly, 
when $\mu < 1$, the points $y = \pm 1$ are still the most probable values
of the magnetization. This comes from the fact that for long flip times,
the magnetization has time to converge towards its asymptotic value. 
However, when the flips become too frequent (i.e. for $\mu > 1$), the
most probable value of the magnetization is zero. For large $\mu$, the 
distribution actually becomes Gaussian, with a width equal to $1/\sqrt{\mu}$.

\section{Effective hyperfine temperature in Gd$_2$Sn$_2$O$_7$}
\label{hypeff}

\begin{figure}
\centerline{
\epsfxsize = 8 cm
\epsfbox{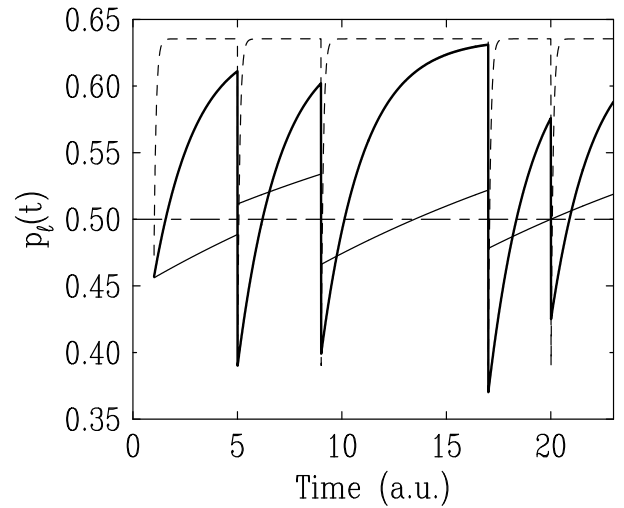}
}
\vskip 0.5 cm
\caption{\sl Temporal evolution of the population $p_\ell(t)$ of the ground 
level of a spin 1/2 doublet as the
hyperfine field reverses randomly in time, for three
values of the ratio $T_1/\tau$. The doublet splitting is $\Delta$=14\,mK and
the sample temperature $T$=33\,mK. 
The instants of field reversal were arbitrarily chosen to occur at
5, 9, 17 and 20\,a.u and the nuclear relaxation has an exponential form.
{\bf Dashed line}: $T_1 \ll \tau$; the hyperfine levels have time 
to thermalize between spin flips, and the time average of $p_\ell$ is the 
Boltzmann value $p_\ell^B$ (0.63); {\bf thick solid line}: 
$T_1 \sim \tau$; the time 
average of $p_\ell$ is smaller than $p_\ell^B$ and it 
is a function of the ration $T_1/\tau$; {\bf thin solid line}:
$T_1 \gg \tau$; the time average of $p_\ell$ tends to 0.50.}
\label{p1t}
\end{figure}

In order to compare with experiment (section \ref{moss}), it is convenient to 
express the probability distribution found above in terms of an effective 
temperature. First, we present in
Fig.\ref{p1t} a schematic representation of the time dependence of the 
population of the lowest energy level for the two limiting cases
of rapid and slow nuclear relaxation with respect to the spin flip, and for
the intermediate case where the two time scales $T_1$ and $\tau$ are of 
comparable magnitude. When $\tau$ is comparable or larger than $T_1$, the
time-averaged population is reduced compared to the Boltzmann value.
In this context it is 
important to emphasize the difference between spin states and energy levels. 
When the 
hyperfine field flips, the spin states exchange their energies, so that the
two spin directions are
equiprobable on average. This is not the case for the energy levels, as
the level with the lowest energy remains the most probable on average.
We now calculate the stationary population of the lowest level. 
We can identify a spin state and an energy level only for a given value of the 
field $H(t)$. For instance, if $H(t)=+H_0$, the lowest energy level 
corresponds to the ``up'' state. The probability $p_\ell^+$ to be in the 
lowest state when $H(t)=+H_0$ is therefore given by:
\be 
p_\ell^+ = \frac{1}{2} (1 + M^+) = \frac{1}{2} (1 + y \frac{M_0}{m_0}).
\ee
In a similar way, the probability to be in the lowest state when the field is 
$-H_0$ is given by:
\be 
p_\ell^- = \frac{1}{2} (1 - M^-) = \frac{1}{2} (1 - y \frac{M_0}{m_0}).
\ee
Therefore, the average population in the lowest level is obtained as:
\be
\langle p_\ell \rangle = \frac{1}{2} + \frac{1}{4} \frac{M_0}{m_0}
\int_{-1}^{1} \left[ y  P_+(y) - y P_-(y) \right]\, dy.
\ee
Using the fact that $P_+(y) = P_-(-y)$, this expression can be transformed into:
\be
\langle p_\ell \rangle = \frac{1}{2} \left[1+ \frac{M_0}{m_0}
\int_{-1}^{1} y  P_+(y) \, dy \right].
\ee

Using the expression of $P_+(y)$ given by Eqn.\ref{pplus}, we find:
\bea
\langle y \rangle_{P_+} &=& \frac{\Gamma(\frac{1}{2}+\mu)}{\Gamma(\frac{1}{2}) 
\Gamma(\mu)}\, \frac{\Gamma(\frac{3}{2}) \Gamma(\mu)}{\Gamma(\frac{3}{2}+\mu)}
\\
	&=& \frac{1}{1+2\mu},
\label{redmu}
\eea
so that finally
\be
\langle p_\ell \rangle(\mu,T) = \frac{1}{2}\,\left(1\,+\,\frac{1}{1+2\mu} 
\tanh\frac{m_0 H_0}{k_BT}\right).
\ee
It is natural to define an effective temperature $T_{eff}$ such that:
\be
\langle p_\ell \rangle(\mu,T) = \frac{1}{1+\exp(-\Delta/k_BT_{eff})},
\ee
where $\Delta=2 m_0 H_0$ is the mean hyperfine splitting.
Therefore, the following relation holds between $T_{eff}$, $T$ and $\mu$:
\be
\tanh{\Delta \over {2k_BT_{eff}}} = \frac{1}{1+2\mu} \tanh{\Delta
\over {2k_BT}}.
\label{deft}
\ee
In the limit $T > 2 \Delta$, which is the temperature range of the
experiments, the above relation reduces to a simple linear function:
\be
T_{eff} \simeq 2T\mu\,+\,T.
\ee
In the other limit, when $T$ goes to zero, we get from Eqn.\ref{deft}:
\be
\tanh{\Delta \over {2k_BT_{eff}}} \to \frac{1}{1+2\mu}.
\ee
So, if $\mu$ does not tend to zero with $T$, $T_{eff}$ remains finite, or even 
goes to infinity if $\mu$ diverges as $T \to$0. As will be discussed in
section \ref{hypspe},
this latter case is plausible because $T_1$ should increase and $\tau$ can
remain finite as $T$ tends to zero. The out of equilibrium two-level system
should then show equipopulation at zero temperature. We note that the concept 
of an effective temperature in out-of-equilibrium systems has been recently
discussed in \cite{CuKu}, where the general situation of a continous degree of
freedom interacting with a multi-time scale random environment is considered.

\begin{figure}
\centerline{
\epsfxsize = 8.5 cm
\epsfbox{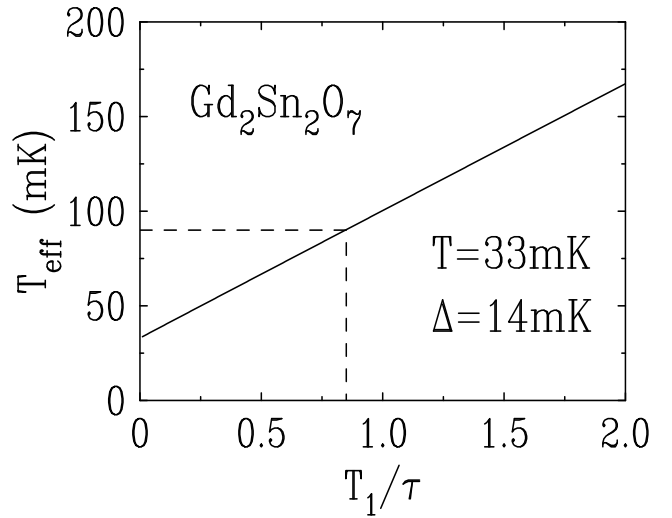}
}
\vskip 0.5 cm
\caption{\sl Effective hyperfine temperature $T_{eff}$ for a spin 1/2 system, 
with a splitting $\Delta$=14\,mK and for a temperature $T=33$\,mK, as 
a function of the ratio of the nuclear relaxation time $T_1$ to the electronic
spin flip time $\tau$. The dashed lines link the effective hyperfine
temperature measured in Gd$_2$Sn$_2$O$_7$ to the $T_1/\tau$ value.}
\label{teff}
\end{figure}

The effective temperature as a function of $\mu=T_1/\tau$
is represented in Fig.\ref{teff} for Gd$_2$Sn$_2$O$_7$, assuming a
sample temperature of 33\,mK and a nuclear splitting of 
14\,mK. The time scale ratio can be read off on the
figure from the value of the measured hyperfine temperature $T_{eff}$; 
choosing the mean value $T_{eff}$=90\,mK,
we find: $T_1/\tau \sim$0.85. Because the experimental accuracy
with which $T_{eff}$ can be measured decreases progressively as the temperature
is increased, it is not possible to obtain reliable values of $T_{eff}$ as a
function of $T$, nor to obtain any information concerning the thermal
dependence of the ratio $\mu$. 

Therefore, in the ground state of Gd$_2$Sn$_2$O$_7$ at very low temperatures, 
the Gd$^{3+}$ magnetic moments are correlated and they fluctuate with a 
characteristic time which is close the nuclear relaxation time. 
The fact that $T_1$ and $\tau$ are of the same magnitude is
somewhat surprising, as the nuclear relaxation times at very low temperature 
in insulators are in the majority of cases quite long ($\sim 10^{-2} - 1$\,s),
whereas the
electronic fluctuation times are rather in the $\mu$s range or shorter.
However, in magnetically ordered (or short range correlated) materials at low 
temperature, the dominant hyperfine relaxation mechanism is the scattering of 
magnons by nuclear moments via the
transverse part of the hyperfine interaction \cite{moriya}. This mechanism
yields nuclear relaxation frequencies proportional to the square of the
hyperfine constant $A$, and can lead 
to short relaxation times, as measured for instance in antiferromagnetic
CrCl$_3$ \cite{narath} where $T_1$ of $^{53}$Cr is in the range 
$10^{-4} - 10^{-3}$\,s in zero or low field at low temperature. As the
hyperfine constant of $^{155}$Gd is about 10 times bigger than that of
$^{53}$Cr, it is likely that the hyperfine relaxation frequencies
fall in the MHz range in Gd$_2$Sn$_2$O$_7$.
Further information concerning the fluctuation rate of the Gd$^{3+}$ spins 
should be provided by our planned muon spin relaxation ($\mu$SR) measurements.

Let us remark that for paramagnetic electronic spins experiencing
an exchange interaction,
the fluctuations would occur at a rate ${1 \over \tau} \sim 
{{k_B \vert \theta_p \vert} \over h}$; taking $\theta_p \sim-$8\,K, 
(the experimental value for Gd$_2$Sn$_2$O$_7$) then 
${1 \over \tau} \sim$100\,GHz. Such fluctuation rates are well above 
the upper limit of the $^{155}$Gd M\"ossbauer ``relaxation window'' 
($\sim$ 1\,GHz) and they would completely wipe out the hyperfine structure.
As the low temperature fluctuations of the Gd$^{3+}$ moments
occur at frequencies below  100\,MHz, they cannot be linked to paramagnetic 
relaxation. This confirms that the fluctuations concern Gd$^{3+}$ 
moments which are correlated.

\section{Influence of the electronic fluctuations on the hyperfine
specific heat}
\label{hypspe}
We will compute here the stationary out-of-equilibrium hyperfine specific
heat when electronic fluctuations are present.
As a function of the ratio $\mu$ of the nuclear relaxation time to the 
electronic flip time, the average stationary populations 
$\langle p_\ell \rangle(\mu,T)$ and $1-\langle p_\ell \rangle(\mu,T)$ of the 
two levels fulfil the relation:
\be
2\langle p_\ell \rangle(\mu,T)-1= g(\mu) (2p_\ell^B - 1),
\ee
where $p_\ell^B$ is the Boltzmann population and $g(\mu)$ a
reduction function given by Eqn.\ref{redmu}:
\be
g(\mu) = \frac{1}{1+2\mu}.
\ee
The energy of the out-of-equilibrium two-level system being given by:
\bea
E(\mu,T) &=& \langle p_\ell \rangle(\mu,T) E_1 \,+\,[1-\langle p_\ell \rangle(\mu,T)] E_2 \nn \\ 
         &=& E_2\,-\,\Delta \langle p_\ell \rangle(\mu,T), 
\eea
the specific heat is readily obtained as:
\be
C_p = g(\mu) C_p^{nuc} - \Delta (p_\ell^B - {1 \over 2}) \, {{d\mu} 
\over {dT}} {{dg} \over {d\mu}},
\label{cpr}
\ee
where $C_p^{nuc}$ is the standard (static) two-level Schottky expression:
\be
C_p^{nuc} = k_B \left(\frac{\Delta}{k_BT}\right)^2 \frac{\exp(-\Delta/k_BT)}
{[1+\exp(-\Delta/k_BT)]^2}.
\ee
The first term in Eqn.\ref{cpr} leads, for finite $\mu$, to a reduction of
the nuclear specific heat with respect to the standard (static) Schottky 
result, since
$g(\mu) < 1$. The sign of the second term depends on the sign of ${{d\mu} \over
{dT}}$ which is difficult to estimate because, in principle, both $T_1$ and
$\tau$ should decrease as temperature increases. A reasonable
assumption is that $T_1$, which is due to a thermally driven mechanism,
varies with temperature more rapidly than $\tau$, which is probably 
associated to a (tunneling) quantum mechanism. Then ${{d\mu} \over {dT}} < 0$,
and the second term in the expression of $C_p$ leads to a further, but
temperature dependent, reduction. Therefore, the specific heat
associated with the hyperfine levels
could be strongly reduced in the presence of electronic fluctuations. 
In particular, when the electronic fluctuations are very fast,
i.e. when the ratio $\mu \gg 1$, the hyperfine specific heat vanishes, 
as both $g$ and ${{dg} \over {d\mu}}$ tend to zero when $\mu$ increases. 

An issue of current interest in geometrically frustrated compounds 
concerns the possible existence of a missing electronic entropy
\cite{ramirez99}. The electronic entropy being deduced from the measured
specific heat, it is necessary to incorporate a correct 
assessment of the contribution of the hyperfine levels, as discussed in 
Ref.\onlinecite{bramwell01}. As shown here, the stationary out-of-equilibrium
hyperfine specific heat appropriate for fluctuating spin systems may 
be smaller than the hyperfine specific heat due to the standard (static)
Schottky anomaly. In such a case, the value of the electronic entropy 
obtained by subtracting the standard (static) Schottky anomaly contribution
from the total measured value would be underestimated.

In theory, all geometrically frustrated systems are potential candidates for a 
reduced hyperfine specific heat because such systems share the common property
of persisting low temperature spin fluctuations. In practice, the amount of
reduction (from no reduction to total removal) will depend on the ratio $\mu$ 
of the nuclear relaxation and electronic spin flip times.
For the rare-earth pyrochlores quite different behaviours are observed.
As shown here, in Gd$_2$Sn$_2$O$_7$ where $\mu \sim$0.85, the hyperfine 
specific heat will be 
reduced at least by a factor 0.37. In Yb$_2$Ti$_2$O$_7$, where the low 
temperature spin flip time is $\sim$ 1\,$\mu$s
\cite{hodg1}, the hyperfine specific heat appears to be present with a value 
approaching its full possible size
\cite{blote69}. An essentially complete hyperfine specific heat appears also 
to be present in Ho$_2$Ti$_2$O$_7$ 
\cite{bramwell01}. In Dy$_2$Ti$_2$O$_7$ however, the hyperfine specific heat
appears to be considerably reduced: the Schottky anomaly calculated using the 
known hyperfine Dy$^{3+}$ hyperfine parameters \cite{almog73}
is not present in the data of Ref.\onlinecite{ramirez99}.
This absence of a hyperfine specific heat provides evidence that the electronic
spin flips persist as T $\to$ 0 in 
Dy$_2$Ti$_2$O$_7$.
Considering the different pyrochlores relative to Fig.\ref{teff}, 
Yb$_2$Ti$_2$O$_7$, Ho$_2$Ti$_2$O$_7$ and Gd$_2$Ti$_2$O$_7$ are situated 
near the left hand side, Gd$_2$Sn$_2$O$_7$ is situated near the centre and
Dy$_2$Ti$_2$O$_7$ is situated near or beyond the right hand side.

\section{Conclusion}

Collective electronic fluctuations have been evidenced by $^{155}$Gd 
M\"ossbauer spectroscopy in the frustrated antiferromagnet  pyrochlore 
Gd$_2$Sn$_2$O$_7$ at very low temperature (27\,mK). These fluctuations show up 
in an unusual manner, i.e. through the out of thermal equilibrium 
populations of the $^{155}$Gd hyperfine levels. We developed a model 
calculation which provides the stationary
populations of a spin 1/2 system in the presence of a 
magnetic field which randomly reverses in time. With simple assumptions, we 
obtain an analytical
expression for the level populations, or equivalently for the effective
temperature of the system, as a function of the ratio of the nuclear relaxation
time to the electronic flip time. Applying this calculation to the case of
Gd$_2$Sn$_2$O$_7$, we find that the ratio of these two time scales is close to
unity. 

From an analogous study of Gd$_2$Ti$_2$O$_7$ we show that the hyperfine levels
are populated at or close to thermal equilibrium. This difference 
compared to Gd$_2$Sn$_2$O$_7$ is probably not due to any difference 
in their basic properties: in both compounds, we find that
the antiferromagnetic  Gd$^{3+}$ - Gd$^{3+}$ coupling has comparable strength,
that there are two magnetic transitions near 1\,K and that, at lower 
temperature, the electronic moments are blocked (on the scale of 
$\sim$ 100\,MHz) along directions perpendicular to the local [111] axes. 
Rather, we suggest the hyperfine populations in Gd$_2$Ti$_2$O$_7$ are at 
thermal equilibrium because the electronic spin fluctuations are partially
quenched under the influence of disorder. 
The presence of disorder in Gd$_2$Ti$_2$O$_7$ is suggested by the abnormally 
enhanced background signal in the X-ray diffraction patterns for
our and other \cite{knop69} samples. In Gd$_2$Sn$_2$O$_7$ there is no
abnormal background signal. The suggestion that the spin fluctuations are 
partially quenched in Gd$_2$Ti$_2$O$_7$ is coherent with the
observation of magnetic Bragg peaks
\cite{champion}. It would be interesting to examine if such peaks are as well
developed in Gd$_2$Sn$_2$O$_7$.

Finally, we examine the consequence of electronic spin fluctuations on the 
hyperfine specific heat and show that it can be strongly reduced when the 
ratio of the nuclear relaxation time to the electronic flip time is comparable 
or greater than unity.

\acknowledgements
We thank J.F. Lericque and N. Genand-Riondet for technical assistance and
A. Forget for preparing the samples.

\end{multicols}

\end{document}